\documentstyle[chicago]{article}

\newcommand{\cov}{{\rm Cov}}
\newcommand{\var}{{\rm Var}}
\newcommand{\ex}{{\rm E}}

\newcommand{\trace}{{\rm Tr}}

\newcommand{\trans}{\mbox{}^{\rm T}}

\newcommand{\tprod}{\otimes}

\newcommand{\inv}{^{-1}}

\newcommand{\reals}{{\rm{I\hspace{-3pt} R}}}
\newcommand{\naturals}{{\rm{I\hspace{-3pt} N}}}
\newcommand{\be}{\begin{eqnarray}}
\newcommand{\ee}{\end{eqnarray}}
\newcommand{\beq}{\begin{equation}}
\newcommand{\eeq}{\end{equation}}
\newcommand{\bd}{{\rm [B/D] }}

\newcommand{\bvec}[1]{\mbox{\boldmath $#1$}}
\newcommand{\mvec}{\mbox{\bf vec}}
\newcommand{\proof}{

\noindent {\it Proof\/}

}
\newcommand{\proofend}{\hfill $\Box$

\vspace{0.2in}

}

\newtheorem{thm}{Theorem}
\newtheorem{lemma}[thm]{Lemma}

\newtheorem{defn}[thm]{Definition}

\title{Bayes linear covariance matrix adjustment for multivariate 
dynamic linear models} 
\author{
Darren J Wilkinson
\thanks{E-mail: {\tt d.j.wilkinson@durham.ac.uk} --- WWW: {\tt
    http://fourier.dur.ac.uk:8000/djw.html} ---
 For information on [B/D], go to {\tt http://fourier.dur.ac.uk:8000/stats/bd/}}
\and
Michael Goldstein \\
}
\date{University of Durham, UK}

\begin{document}
\maketitle

\begin{abstract}
A methodology is developed for the adjustment of the
covariance matrices underlying a multivariate constant time series
dynamic linear
model. The covariance matrices are embedded in a distribution-free 
inner-product space of matrix objects which facilitates such adjustment.
This approach helps to make the analysis simple, tractable and
robust. 
To illustrate the methods, a simple model is 
developed for a time
series representing sales of certain brands of a product from a 
cash-and-carry depot. The covariance structure underlying the model is
revised, and the benefits of this revision
on first order inferences are then examined.
\end{abstract}


\section{Introduction}
Bayesian covariance matrix estimation is a notoriously difficult problem. For
large matrices in particular, specification of beliefs about the
distribution of the matrix is particularly hard. Distributional
Bayesian approaches to the problem have tended to make use of restrictive
conjugate prior assumptions in order to fix the prior distribution with as few
hyper-parameters as possible. More recently, \citeN{tlcovmat} have
weakened the distributional assumptions required. However,
specification within their formulation is very difficult, and, for
large matrices, the computational problems are considerable.
Specification is somewhat easier using the approach of \citeN{blzicm},
but the other problems remain.
In \citeN{djwgvar}, we outline a new
approach to covariance matrix adjustment, exploiting second order
exchangeability specifications, and a geometric space where random
matrices live naturally.

Covariance matrix adjustment for dynamic linear models is reviewed
in \citeN{dlm}. For multivariate time series, the
observational covariance matrix can be updated for a class of models
known as {\it matrix normal models\/} using a simple conjugate prior approach.
However, the distributional assumptions required are extremely restrictive,
and it is difficult to learn about the covariance matrix for the 
updating of the state vector.

In this paper, we apply our approach to
develop a
methodology for the revision of the underlying covariance
structures
for a
dynamic linear model,
free from any distributional restrictions, 
 using Bayes linear estimators for the
covariance matrices based on simple quadratic observables. We do this
by constructing an inner-product space of random
matrices containing both the underlying covariance matrices and observables
predictive for them. Bayes linear estimates for the
underlying matrices follow by orthogonal projection.

We illustrate the method with data derived from the weekly sales of 
six leading brands of shampoo  from a
medium sized cash-and-carry depot. The sales are
modelled taking into account the 
underlying demand and competition effects, and the covariance
structure over the resulting dynamic linear model
is adjusted using the weekly sales data. 

\section{The dynamic linear model}

\subsection{The general model}

\label{cdlm}
Let $\bvec{X}_1,\bvec{X}_2,\ldots$ be an infinite sequence
of random vectors, each of length $r$, such that $\bvec{X}_t=
(X_{1t},X_{2t},\ldots,X_{rt})\trans$. These vectors represent the
observations at each time point. Suppose that we model the relationships
between these vectors in the following way.
\be
\bvec{X}_t &=& F\trans\bvec{\Theta}_t + \bvec{\nu}_t \\
\bvec{\Theta}_t &=& G\bvec{\Theta}_{t-1} + \bvec{\omega}_t
\ee
Our prior second-order specification is as follows: 
\beq
\ex(\bvec{\nu}_t)=\ex(\bvec{\omega}_t)=0,
\var(\bvec{\Theta}_0)=\Sigma, \var(\bvec{\nu}_t)=V,
\var(\bvec{\omega}_t)=W,\quad\forall t
\label{spec1}
\eeq
\beq 
\cov(\bvec{\Theta}_s,\bvec{\nu}_t)=\cov(\bvec{\Theta}_s,\bvec{\omega}_t)
=\cov(\bvec{\nu}_s,\bvec{\omega}_t)=0\quad\forall s,t,\ 
\cov(\bvec{\omega}_s,\bvec{\omega}_t)=
\cov(\bvec{\nu}_s,\bvec{\nu}_t)=0\quad\forall s\not=t
\label{spec2}
\eeq
where we adopt the 
usual conventions
$\cov(\bvec{A},\bvec{B})=\ex(\bvec{AB}\trans)-
\ex(\bvec{A})\ex(\bvec{B}\trans)
$
 and
$\var(\bvec{A})=\cov(\bvec{A},\bvec{A})$.
The {\em state vector\/}, $\bvec{\Theta}_t$ is $p$ dimensional, and the
$p\times r$ and $p\times p$ dimensional matrices, $F$ and $G$ are 
assumed to be known. 
This is a second-order description of the (constant) multivariate
time series dynamic linear model (DLM) described in
\citeN{dlm}. We make no
distributional assumptions for any of the components in the model.
In this paper, we describe ways to learn about $V$ and $W$ from data.
\citeN{dlm} (Chapter 15) give a conjugate prior solution to the 
problem of learning
about $V$ for a class of these models known as {\em matrix normal
  models\/}, if one is prepared to make the necessary distributional
assumptions. However methods for learning about the
matrix $W$ tend primarily to be {\em ad hoc\/}.

\subsection{Example}
To illustrate our approach, we consider 
a simple locally constant model for the sales of 6 leading
brands of shampoo from
a medium sized cash-and-carry depot.
As above $\bvec{X}_1,\bvec{X}_2,\ldots$ is our sequence
of random vectors, each of length $6$, such that $\bvec{X}_t=
(X_{1t},X_{2t},\ldots,X_{6t})\trans$.
 The component $X_{it}$ represents the (unknown) sales of brand $i$ at
 time $t$. The vectors of sales are modelled as follows
\beq
\bvec{X}_{t} = \bvec{\Theta}_{t} + \bvec{\nu}_{t} \quad
\forall t
\label{exobs}
\eeq
where
\beq
\bvec{\Theta}_{t} = \bvec{\Theta}_{t-1} + \bvec{\omega}_t \quad \forall t
\label{extheta}
\eeq
Prior beliefs are are given by (\ref{spec1}) and (\ref{spec2}).
Here we are assuming the process to be locally constant, but
with different underlying demands for each of the components of the
series. This is a simple model, with no seasonal component, chosen 
to illustrate our
methodology, and
would be unrealistic if there were noticeable trends within any of the
components of the series. However, for high dimensional time series
with no obvious trends, it is often
the case that, provided the covariance structure is appropriate, many of the
interesting features of the series can be captured using just such a
model. To this end we
introduce covariances between components of the state vector and also for the
way demand changes over time, and for the way observations vary from
the underlying demand.
A more detailed treatment of multivariate
sales forecasting within
a fully specified Bayesian framework is given by \citeN{cmqbayes} 
and \citeN{cmqcomp} who consider
the problem of developing a dynamic  model for 
multivariate sales, and the development of a prior distribution with
sufficient flexibility to capture the effects of market interaction.

The second-order DLM requires the following quantifications.
 Firstly, the $F$ and $G$
matrices must be specified. Then, {\it
  a priori\/} specifications are needed for the expectation of the
initial state vector, $\bvec{\mu}_0=\ex(\bvec{\Theta}_0)$.
Finally, we must
specify the matrices $\Sigma=\var(\bvec{\Theta}_0),\
V=\var(\bvec{\nu}_t),\ W= \var(\bvec{\omega}_t)\forall t$.

In our example the specification for the
mean vector was
\beq
\ex(\bvec{\Theta}_0)=(10,9,9,8,8,7)\trans
\label{thetaspec}
\eeq
The following specifications were made for the covariance matrices,
using exchangeability judgements concerning way the observations
vary from their means. Using the notation $A_{ij}$ for the
$(i,j)^{th}$ element of the matrix $A$, we have
\be 
\Sigma_{ii}&=&9\ \forall i,\ 
\Sigma_{ij}=3\ \forall i\not=j,
\label{varsigma}
\\
W_{ii}&=&4\ \forall i,\ 
W_{ij}=1\ \forall i\not=j
\label{varomega}
\\
V_{ii}&=&36\ \forall i,\ 
V_{ij}=-4\ \forall i\not=j
\label{varnu}
\ee
In truth, there is perhaps more symmetry in these specifications than is
really appropriate, but specification is hard, and viewing variation
in the sales of
the various shampoos as second-order exchangeable 
 greatly reduces the number of
specifications
which we have to make over the second order structure, and will
allow further exchangeability modelling to simplify the fourth order
specifications in later sections.

Notice however that many aspects of the underlying
mechanisms have been captured by these specifications.
In this model, $\bvec{\Theta}_t$ represents the vector of
demands at time $t$. From the 
positive correlations in $\var(\bvec{\Theta}_0)$, if the mean of one
product turned out to be higher than anticipated, we would revise upwards
beliefs about the means of the other products. Also, the positive correlations
within $\var(\bvec{\omega}_t)$ indicate that there is a common component
to the demands, whilst the negative correlations within $\var(\bvec{\nu}_{t})$
indicate that brands are competing, and tend to succeed at the expense
of each other.

\subsection{Bayes linear analysis}
We take a Bayes linear approach to subjective statistical inference, making
expectation (rather than probability) primitive. An overview of the methodology
is given in \citeN{fgcross}, in which the emphasis is on learning
about means. 
With the second-order specification that we have made, we may use
sales data to
carry out an analagous Bayes linear analysis
which will be informative for the mean of future observations.
 However, we will not
learn about the covariance matrices
$W=\var(\bvec{\omega}_t)$ or $V=\var(\bvec{\nu}_t)$. In this paper, we
 describe how such learning may take place.

\section{Quadratic products}
\subsection{Exchangeable decomposition of unobservable products}
For the matrix
$A=(\bvec{a}_1,\bvec{a}_2,\ldots,\bvec{a}_n)$, we define
\beq
\mvec A = (\bvec{a}_1\trans,\bvec{a}_2\trans,\ldots,\bvec{a}_n\trans)\trans
\eeq
For the general DLM outlined in Section \ref{cdlm}, we
form the quadratic products of $\bvec{\omega}_{t}$ and
  $\bvec{\nu}_{t}$, namely $\mvec(\bvec{\omega}_t\bvec{\omega}_t\trans)$ and
$\mvec(\bvec{\nu}_t\bvec{\nu}_t\trans)$.
We view $\mvec(\bvec{\omega}_t\bvec{\omega}_t\trans)$ and 
$\mvec(\bvec{\nu}_t\bvec{\nu}_t\trans)$ to be
second-order exchangeable over $t$. By this, we mean that our
second-order beliefs
over the vectors of quadratic products of residuals will remain 
invariant under the action of an
arbitrary permutation of the $t$ index
 (this is 
what we mean
when describing a DLM as {\em constant\/}). From
the second-order exchangeability representation
theorem \cite{mgexchbel}, we may represent an element of a
second-order exchangeable collection of vectors as the sum of a mean
vector, common to all elements,
and a residual vector, uncorrelated with the mean vector and
all other residual vectors. We may apply this representation to
$\mvec(\bvec{\omega}_t\bvec{\omega}_t\trans)$, and then re-write the
representation in matrix form as
\beq
\bvec{\omega}_{t}\bvec{\omega}_{t}\trans = V^{\omega} +
S^{\omega}_{t}\quad
 \forall t\geq 1
\label{omegarep}
\eeq
where $V^\omega$ and $S^\omega_t$ are random matrices of the same dimension as
$\bvec{\omega}_{t}\bvec{\omega}_{t}\trans$,
$\ex(\mvec(S^\omega_{t}))=0$ and $\cov(\mvec(V^\omega),\linebreak
\mvec(S^\omega_t))= 
\cov(\mvec(S^{\omega}_{t}),\mvec(S^{\omega}_{s}))=0
,\ \forall s\not= t$, $\var(\mvec S_s^\omega)=\var(\mvec S_t^\omega),\
\forall s,t$.
Decomposing $\mvec(\bvec{\nu}_t\bvec{\nu}_t\trans)$ similarly, we obtain
\beq
\bvec{\nu}_{t}\bvec{\nu}_{t}\trans = V^\nu + S^\nu_{t}\quad \forall t\geq 1
\label{nurep}
\eeq
with properties as for representation (\ref{omegarep}).
Note that $\ex(V^\omega)=\ex(\bvec{\omega}_t\bvec{\omega}_{t}\trans)=
\var(\bvec{\omega}_{t})=W$ and 
so learning about $V^\omega$ will allow us to learn about the covariance matrix
for the residuals for the state, and
$\ex(V^\nu)=\ex(\bvec{\nu}_{t}\bvec{\nu}_{t}\trans)=\var(\bvec{\nu}_{t})=V$,
and so learning about $V^\nu$ will allow us to learn about the
covariance
matrix 
for the observational residuals. Representations (\ref{omegarep}) and
(\ref{nurep}) decompose our uncertainty about 
$\bvec{\omega}_t\bvec{\omega}_{t}\trans$ and
$\bvec{\nu}_{t}\bvec{\nu}_{t}
\trans$ into two parts. Bayes linear updating (with enough data)
will eliminate the aspects of uncertainty derived from uncertainty
about $V^\omega$ and $V^\nu$.

In order to conduct a Bayes linear analysis on the quadratic
structure we need additional covariance specifications
  $\var(\mvec{V}^\omega)$, $\var(\mvec{V}^\nu)$,
  $\var(\mvec{S}^\omega_{t})$ and 
 $\var(\mvec{S}^\nu_{t})$, for some $t$.

\subsection{Example}
\label{exspecs}
In our example the $\bvec{X}_t$ vector is 6-dimensional, and so the
matrices, 
$V^\omega$, $V^\nu$, $S^\omega$ and $S^\nu$ are $6\times
6$-dimensional. 
Consequently, the
matrices
$\var(\mvec{V}^\omega)$, $\var(\mvec{V}^\nu)$,
  $\var(\mvec{S}^\omega_{t})$ and 
 $\var(\mvec{S}^\nu_{t})$
 are $36\times 36$-dimensional. 
When referring to the
components of $\var(\mvec{V}^\omega)$, the notation $v^\omega_{ijkl}$
 will be used to denote
the covariance between the $(i,j)^{th}$ and $(k,l)^{th}$ elements of
$V^\omega$. Similar notation is used for $\var(\mvec{V}^\nu)$.
Also $s^\omega_{ijkl}$ and $s^\nu_{ijkl}$ are used for the components
of $\var(\mvec{S}^\omega_{t})$ and 
 $\var(\mvec{S}^\nu_{t})$ respectively.
The following covariance specifications were made for our example:
\beq
v^\omega_{iiii}= 9/4,\ \forall i,\quad
v^\omega_{ijij}= 9/16,\ \forall i\not=j,\quad
v^\omega_{iijj}= 1/5, \forall i\not= j,\quad
\label{qspec1}
\eeq
\beq
v^\nu_{iiii}= 25,\ \forall i,\quad
v^\nu_{ijij}= 1,\ \forall i\not=j,\quad
v^\nu_{iijj}= 4,\ \forall i\not=j,\quad
\label{qspec2}
\eeq
\beq
s^\omega_{iiii}= 30,\ \forall i,\quad
s^\omega_{ijij}= 15,\ \forall i\not=j,\quad
s^\nu_{iiii}= 2500,\ \forall i,\quad
s^\nu_{ijij}= 1000,\ \forall i\not=j.
\label{qspec3}
\eeq
For instance, $v^\omega_{iiii}$ is the variance specification for the
$(i,i)^{th}$ element of $V^\omega$, which represents
the underlying variance of the $i^{th}$ element of $\bvec{\omega}_t$.
From (\ref{varomega}), it has expectation 4. From (\ref{extheta}), 
this value governs the rate of change 
of $\bvec{\Theta}_t$. By considering the
range of plausible variances for the way $\bvec{\Theta}_t$ might change
over time, it was felt reasonable that a standard
deviation specification of 3/2 should be made. The
other specifications were made in a similar fashion.
For simplicity
in this example, the specifications for $s^\omega_{ijkl}$ and
$s^\nu_{ijkl}$  were made using the fourth moments of the multivariate
normal distribution compatible with the given second-order structure as
a guide.

While we would require considerably more specifications for a full
Bayes or
Bayes linear analysis, (\ref{qspec1}), (\ref{qspec2}), and
(\ref{qspec3})
  are sufficient for our purposes
as these are the only specifications needed for the matrix object
approach to belief revision which we are to take in the later
sections.

 There is a 
lot of symmetry in these values, greatly simplifying
the specification, but once again, any non-negative covariance structure
over the quadratic products is acceptable. Here we have
used assumptions of exchangeability over the variances and the
covariances.  Many of the specifications 
made for the quadratic structure will be
``averaged over" in the matrix object approach to covariance 
adjustment which we shall develop, and so there is
a limit to the effort that we may wish to put into very detailed
specifications at this stage since our suggested analysis will not be
overly sensitive to the individual specifications.

\subsection{Observable quadratic terms}
\label{oqt}
We will first construct certain linear combinations of the observables
 which do not involve the state vector, $\bvec{\Theta}_t$.
This is useful for various reasons, and in particular
 because it greatly reduces the prior
specification required for the analysis of the quadratic structure.
In this paper, we shall mainly be concerned with DLMs for which
 there exists an $r\times r$ matrix $H$,
such that $HF\trans=F\trans G$.
We call such DLMs {\em two-step
  invertible\/}. Note that a DLM will be two-step invertible if $F$ is
of full rank and $r\geq p$ (as will often be the case for
high-dimensional time series), and there will often be many matrices
$H$ satisfying $HF\trans=F\trans G$. For example, $H=F\trans
G\trans {F\trans}^\dagger$ (where ${F\trans}^\dagger$ represents any
generalised inverse of $F\trans$) is a solution.
 Further, if $F$ is of full rank, $r< p$
 and such a
matrix exists, then $H=F\trans GF(F\trans F)\inv$, and so $H$ exists,
if and only if $F\trans GF(F\trans F)\inv F\trans=F\trans G$. 
Note also that the
matrix $H^2$ has the property that $H^2F\trans=F\trans G^2$.
For a two-step invertible DLM, we may construct the following vectors of
observables which do not involve the state vector:
\be
\bvec{X}_t^{\prime} &=& \bvec{X}_{t} - H\bvec{X}_{t-1} = F\trans
\bvec{\omega}_{t}+\bvec{\nu}_{t}-H\bvec{\nu}_{t-1}\quad \forall t\geq 2\\
\bvec{X}_t^{\prime\prime} &=& \bvec{X}_{t} - H^2\bvec{X}_{t-2} = F\trans
\bvec{\omega}_{t}+ F\trans G
\bvec{\omega}_{t-1}  +\bvec{\nu}_{t}-H^2\bvec{\nu}_{t-2} \quad \forall
t\geq 3
\ee
We form the matrices of quadratic products,
$\bvec{X}_t^{\prime}\bvec{X}_t^{\prime}\trans$ and
$\bvec{X}_t^{\prime\prime}\bvec{X}_t^{\prime\prime}\trans\quad\forall
t$. As we shall see, these are predictive for $V^\omega$ and $V^\nu$.

Not all DLMs are two-step invertible,
but for the constant dynamic linear model outlined in Section
\ref{cdlm},  it is always possible to construct linear combinations of the 
observations which do not involve the state, provided only that the 
constant dynamic linear
model in question is {\em observable\/} (for a discussion of the very
weak restriction of
observability, see \citeN{dlm}, Chapter 5). However, in general such
linear combinations require more than two
successive observations from the series.
Consequently, for simplicity we restrict attention to the two-step
invertible model. However, the approach is quite general, and may be
applied similarly for any constant observable DLM, the only difference
being that the covariance specification is more complicated, and more
quantities are involved in the adjustment.

\subsection{Example}
\label{oqte}
For our example, $F$, $G$ and $H$ are all the identity, and so
 we form the one
and two-step differences of the observables:
\be
\bvec{X}^{(1)}_{t} &=& \bvec{X}_{t} - \bvec{X}_{t-1} = \bvec{\omega}_{t} +
\bvec{\nu}_{t} - \bvec{\nu}_{t-1}\label{xip} \quad \forall t\geq 2\\
\bvec{X}^{(2)}_{t} &=& \bvec{X}_{t} - \bvec{X}_{t-2} =
\bvec{\omega}_{t} + 
\bvec{\omega}_{t-1} \label{xipp}
+ \bvec{\nu}_{t} - \bvec{\nu}_{t-2}\quad \forall t\geq 3
\ee
 We then form the quadratic products of these, 
$\bvec{X}^{(1)}_{t}\bvec{X}^{(1)}_{t}\trans$ and 
$\bvec{X}^{(2)}_{t}\bvec{X}^{(2)}_{t}\trans$. 
 These observables are predictive for $V^\omega$
  and $V^\nu$ and so may be used to learn about
  the underlying covariance structure.
 All means and
covariances that we require for the subsequent analysis are determined
by specifications in (\ref{thetaspec}), (\ref{varsigma}), 
(\ref{varomega}), (\ref{varnu}), (\ref{qspec1}),
(\ref{qspec2})  and
(\ref{qspec3}). The precise form of the covariance structure over the
observables 
is rather complex, and is given in the appendix.

\section{$n$-step exchangeability}
\subsection{$n$-step exchangeable collections}
The covariance structure over
$\bvec{X}^{\prime}_t$, $\bvec{X}^{\prime\prime}_t$  and
their quadratic products has a general structure common amongst
ordered vectors of random quantities of the type arising from
differenced time series. The
covariance structure is invariant under arbitrary
translations and reflections of the ordering, and the
auto-correlation function becomes constant after some distance, $n$. 
We call ordered vectors with this property, {\em second-order $n$-step
  exchangeable\/}.
Covariance may be interpreted as an inner-product on a space of
random quantities. We will find it useful to consider
inner-products on spaces of more general random entities. In
particular, in Section \ref{matrices} we shall define an inner-product
 on a space of
random matrices. We require a concept of $n$-step exchangeability
which is sufficiently general that it is also valid for 
this space of matrices,
and so we formalise the concept as follows.

Let
$\{Y_{jk}|\forall j,k\}$ be a collection of random entities of
interest to us. Also form a maximal linearly independent collection
of constant entities of the same type, and call this
collection $C=[C_1,C_2,\ldots]$. When we are dealing with
random scalars, $C$ will consist of the single scalar, $C_1=1$. In
Section \ref{matrices}, we describe the constant space for a
collection of random matrices.
Form the vector space
\beq
{\cal V}=span\{C_j, Y_{jk} | \forall j,k\}
\eeq
so that the random entities are now vectors within this space.
Define an inner-product $(\cdot,\cdot):{\cal V}\times {\cal V}
 \longrightarrow \reals$
on ${\cal V}$.
The inner-product should capture certain aspects of our beliefs about the
relationships between the elements of ${\cal V}$; for scalars, we
might use $(X,Y)=\ex(XY)$. 
Form the completion of the
space ${\cal V}$, and denote this Hilbert space by ${\cal H}$.
Also define a bounded linear expectation function on the elements of
$\cal V$. For the purposes of this paper it suffices to think in
 terms  of the usual definition of
 expectation. In general, however, we
define a bounded
linear function $\ex(\cdot) : {\cal H} \longrightarrow span\{C\}$, such that 
$\forall Y\in {\cal H},\ \ex(Y)$ is 
the orthogonal projection of $Y$ into $span\{C\}$, with 
respect to the inner-product $(\cdot,\cdot)$.
This
is the generalisation of expectation which we require for general
spaces, and is defined in terms of the inner-product. It coincides with the
usual definitions for all of the examples in this paper.

 If $\exists n\in\naturals$ such that $\ex(Y_{jk})=e_j \quad \forall
 j,k$, and
\be
(Y_{ik},Y_{jl}) &=& d_{0ij} \qquad \forall i,j,|k-l|=0 \nonumber \\
(Y_{ik},Y_{jl}) &=& d_{1ij} \qquad \forall i,j,|k-l|=1 \nonumber \\
\vdots & \vdots & \vdots \qquad\qquad \vdots \nonumber \\
(Y_{ik},Y_{jl}) &=& d_{(n-1)ij} \qquad \forall i,j,|k-l|=n-1 \nonumber \\
(Y_{ik},Y_{jl}) &=& c_{ij} \qquad \forall i,j,|k-l|\geq n
\ee
 the collection $\{Y_{jk}|\forall j,k\}$ is said to 
be {\em generalised second-order $n$-step exchangeable 
over $k$\/}.

We apply this definition to the elements,
$X^{\star}=\{X^{\prime}_{it}X^{\prime}_{jt}|\forall i,j,\forall t\geq
2\}$ and
$X^{\star\star}=\{X^{\prime\prime}_{it}X^{\prime\prime}_{jt}|\forall
i,j,\forall t\geq 3\}$,  of the matrices $\{\bvec{X}_{t}^{\prime}
\bvec{X}_{t}^{\prime}\trans|\forall t\geq 2\}$ and 
$\{\bvec{X}_{t}^{{\prime\prime}}
\bvec{X}_{t}^{{\prime\prime}}\trans|\forall t\geq 3\}$ defined in
Section \ref{oqt}, where $X^{\prime}_{it}$ denotes the $i^{th}$ component of 
the vector $\bvec{X}^{\prime}_t$.
 Form a vector space, ${\cal V}$ consisting of all linear combinations
 of the elements of
$X^{\star}$ and $X^{\star\star}$
and the unit constant, and define the inner-product on this space as
\(
(X,Y) = \ex(XY),\quad \forall X,Y\in {\cal V}
\).
We may easily check that $X^\star$ is (second order)
2-step exchangeable over $t$, and that 
$X^{\star\star}$ is 3-step 
exchangeable over $t$.

\subsection{Representation for $n$-step exchangeable collections}
\citeN{mgexchbel} constructs a general representation for second-order
exchangeable collections. There 
is an analagous
representation for collections with the weaker property
of $n$-step exchangeability, constructed in a similar way.
\begin{thm}
\label{nsethm}
Let $\{ Y_{jk}|\forall j,k\}$ be
generalised second-order $n$-step exchangeable over $k$ with respect
to the inner-product $(\cdot,\cdot)$.
 Then the $Y_{jk}$ may be represented as
\beq
Y_{jk}=M_j+R_{jk} \quad \forall j,k
\label{nserep}
\eeq
where the $M_j$ and $R_{jk}$ have the following properties:
\be
\ex(Y_{jk})&=&\ex(M_j),\ \ex({R}_{jk})=0 \quad\forall j,k\\
(M_i,M_j)&=&(M_i,Y_{jk})=c_{ij},
\quad (M_i,R_{jk})=0 \quad \forall i,j,k\\
(R_{ik},R_{jl}) &=& (Y_{ik},R_{jl}) =
  (Y_{ik},Y_{jl})-c_{ij}\quad\forall i,j,k,l
\ee
Further, 
the $\{R_{jk}|\forall j,k\}$ are generalised second-order $n$-step 
exchangeable over $k$, with $(R_{ik},R_{jl})=0\quad\forall i,j,|k-l|\geq n$.
\end{thm}
\proof
Let
\beq
M_{im}=\frac{1}{m}\sum_{k=1}^m Y_{ik} \quad\forall i,m
\eeq
Observe that the sequence $M_{i1},M_{i2},\ldots$ is {\it
  Cauchy} $\forall i$ ie. that $(M_{ik}-M_{il},M_{ik}-M_{il}) 
\longrightarrow 0$ as $k,l
\longrightarrow \infty$, which follows directly from the properties 
of $n$-step exchangeable
sequences.
Construct the quantity $M_{i}$ to be the Cauchy limit of
this sequence so that 
\beq
\lim_{m\rightarrow\infty}(M_{im},Y) = (M_i,Y)\ \forall i,\forall Y\in
{\cal H}
\label{limit}
\eeq
Linearity of $\ex(\cdot)$ gives $\ex(M_{im})=e_i \quad
\forall i,m$, and hence applying (\ref{limit}) for $Y\in C$
 we deduce $\ex(M_i)=e_i$. 
Define $R_{ik}$ via $R_{ik}=X_{ik}-M_{i}\ \forall i$, so that $\ex(R_{im})=0
\quad \forall i,m$.
The other properties of the representation follow directly from (\ref{limit}).
\proofend
As for the case of second-order exchangeability, the mean components of
the representation, $M_j$, represent the quantities which we may learn
about by linear fitting on the data. We may resolve as much uncertainty as we
wish about these quantities given a sufficient number of observations,
by such linear fitting.
We therefore say that the $n$-step exchangeable collection 
$\{{Y}_{jk}|\forall j,k\}$ with representation
${Y}_{jk}={M}_j+{R}_{jk} \quad \forall j,k$
{\em identify\/} the random quantities ${M}_j,\ \forall j$.

\subsection{Identification of the covariance structure underlying the DLM}
The $n$-step exchangeability representation theorem allows
us to construct models for the observable quadratic products
we have formed. The elements of the collection 
$\{\bvec{X}^{{\prime}}_t\bvec{X}^{{\prime}}_t\trans|\forall
t\geq 2\}$ for the two-step invertible DLM, are 2-step exchangeable
over $t$. From Theorem \ref{nsethm}, we may construct the
representation (\ref{nserep}).
The identified quantities may be constructed as the
Cauchy limit of the arithmetic means of the elements.
\begin{lemma}
The 2-step exchangeable collection 
$\{\bvec{X}^{{\prime}}_t\bvec{X}^{{\prime}}_t\trans|\forall
t\geq 2\}$ identify the matrix
\beq
{M^\prime} = F\trans V^\omega F+V^\nu +HV^\nu H\trans 
\eeq
and the 3-step exchangeable collection 
$\{\bvec{X}^{{\prime\prime}}_t\bvec{X}^{{\prime\prime}}_t\trans|\forall
t\geq 3\}$ identify
\beq
{M}^{\prime\prime} = F\trans GV^\omega G\trans F+ 
 F\trans V^\omega F+V^\nu +H^2V^\nu H^2\trans 
\eeq
\end{lemma}
\proof
\be
{M^\prime}&=& \lim_{N\rightarrow\infty}\frac{1}{N}\sum_{t=2}^N
 \bvec{X}_t^{\prime}\bvec{X}_t^{\prime}\trans 
\\
&=&
\lim_{N\rightarrow\infty}\frac{1}{N}\sum_{t=2}^N
 (F\trans
\bvec{\omega}_{t}+\bvec{\nu}_{t}-H\bvec{\nu}_{t-1})(F\trans
\bvec{\omega}_{t}+\bvec{\nu}_{t}-H\bvec{\nu}_{t-1})\trans\\
&=&\lim_{N\rightarrow\infty}\frac{1}{N}\sum_{t=2}^N
 F\trans\bvec{\omega}_t\bvec{\omega}_t\trans F + 
\bvec{\nu}_t\bvec{\nu}_t\trans+
H\bvec{\nu}_{t-1}\bvec{\nu}_{t-1}\trans H\trans  \\
&=& F\trans V^\omega F+V^\nu +HV^\nu H\trans 
\ee
The derivation of ${M}^{\prime\prime}$ is similar.
\proofend
 Now since
\(
\frac{1}{2}\left[HM^{\prime}H\trans-(M^{\prime\prime}-M^{\prime})
\right] = HV^\nu H\trans
\)
we deduce that for a 2-step invertible series, the collection
\beq
\left\{\left. \frac{1}{2}
\left[H\bvec{X}^{{\prime}}_t\bvec{X}^{{\prime}}_t\trans
 H\trans-(\bvec{X}^{{\prime\prime}}_t\bvec{X}^{{\prime\prime}}_t\trans-
\bvec{X}^{{\prime}}_t\bvec{X}^{{\prime}}_t\trans)
\right] \right| \forall t\geq 3 \right\}
\eeq
identifies $ HV^\nu H\trans$.

Now if $r\geq p$, $H$ can usually be chosen to be invertible. If $r<p$,
since we assume that $F$ and $G$ are of full rank, so is $H$, and so $H$ is
invertible. In this paper we restrict attention to those two-step
invertible DLMs which have an invertible $H$. Consequently, since
\(
\frac{1}{2}[M^{\prime}-H\inv(M^{\prime\prime}-M^{\prime}) 
H\inv\trans ] = V^\nu
\)
and
\(
M^\prime-V^\nu-HV^\nu H\trans =
F\trans V^\omega F
\)
we get
\begin{thm}
\label{idquant}
For a 2-step invertible series with invertible $H$,
put 
\beq
{M}_t =
\frac{1}{2}\left[\bvec{X}^{{\prime}}_t\bvec{X}^{{\prime}}_t\trans-
H\inv(\bvec{X}^{{\prime\prime}}_t\bvec{X}^{{\prime\prime}}_t\trans-
\bvec{X}^{{\prime}}_t\bvec{X}^{{\prime}}_t\trans)H\inv\trans
\right]
\eeq

\noindent (i) The collection of matrices
$
\left\{\left. {M}_t  \right| \forall t\geq 3 \right\}
$
identify $ V^\nu$.

\noindent (ii) The collection of matrices
$
\left\{\left.
 \bvec{X}^{{\prime}}_t\bvec{X}^{{\prime}}_t\trans - M_t -
HM_t H\trans 
\right| \forall t\geq 3 \right\}
$
 identify
$F\trans V^\omega F$.
\end{thm}
\proofend
If $p>r$, we do not identify the entire matrix $V^\omega$.
This would require a fuller selection of observables than we have
considered here. The identified matrix, $F\trans V^\omega F$, is the
contribution to the uncertainty for $\bvec{X}_t$ from
$\bvec{\Theta}_t$, given $\bvec{\Theta}_{t-1}$.

\subsection{Example}
In our example, Theorem \ref{idquant} implies that
 the collection $\{\bvec{X}^{(2)}_t\bvec{X}^{(2)}_t\trans
  - \bvec{X}_t^{{(1)}}\bvec{X}_t^{{(1)}}\trans | \forall t\geq 3 \}$
  identify $ V^\omega$ and that
  the collection $\{\bvec{X}_t^{{(1)}}\bvec{X}_t^{{(1)}}\trans
  - \frac{1}{2}\bvec{X}_t^{(2)}\bvec{X}_t^{(2)}\trans | \forall
  t\geq 3 \}$
identify $ V^\nu$.

By observing sales at an increasing (but finite)
number of time points, we may resolve through linear fitting,
 as much uncertainty
as we wish about the underlying covariance structure 
for the particular 
time series model we are dealing with.

If all fourth order prior belief specifications have been made, a
simple Bayes 
linear analysis can be carried
  out in order to learn about the underlying covariance structure by 
adjusting the elements of $ V^\nu,\  V^\omega$ by the
elements of the observable matrices
$\bvec{X}_t^{{(1)}}\bvec{X}_t^{{(1)}}\trans,\ 
\bvec{X}_t^{(2)}\bvec{X}_t^{(2)}\trans$. However,
for long, high-dimensional time series, the 
number of quantities involved in a full linear adjustment
is extremely 
large, and so it is important to reduce the dimensionality of
the problem, and  preserve the
inherent matrix structure.

\section{Matrix objects for the time series}
\subsection{Formation of the matrix space}
\label{matrices}
We are interested in learning about $r\times r$ dimensional covariance
matrices. We first form a
collection of $r^2$ linearly independent $r\times r$ constant matrices
such that
$C_{r(i-1)+j}$ is the matrix with a $1$ in the $(i,j)^{th}$ position, and 
zeros elsewhere, where $i$ and $j$ range from $1$ to $r$ and call this
collection $C=
[C_1,\ldots,C_{r^2}]$.
 This collection of matrices is a basis for
the space of {\it known\/} $r\times r$ matrices. Define the
collections of matrices
\beq
X_2^\ddagger=\{\bvec{X}_2^{{\prime}}\bvec{X}_2^{{\prime}}\trans,
 H\bvec{X}_2^{{\prime}}\bvec{X}_2^{{\prime}}\trans H\trans, 
H\inv \bvec{X}_2^{{\prime}}\bvec{X}_2^{{\prime}}\trans H\inv\trans \}
\eeq
\beq
X_t^\ddagger=\{\bvec{X}_t^{{\prime}}\bvec{X}_t^{{\prime}}\trans,
\bvec{X}_t^{{\prime\prime}}\bvec{X}_t^{{\prime\prime}}\trans,
 H\bvec{X}_t^{{\prime}}\bvec{X}_t^{{\prime}}\trans H\trans, 
H\inv \bvec{X}_t^{{\prime}}\bvec{X}_t^{{\prime}}\trans H\inv\trans , 
H\inv \bvec{X}_t^{{\prime\prime}}\bvec{X}_t^{{\prime\prime}}\trans H\inv\trans 
\},\quad \forall t\geq 3
\eeq
Form the real vector space, ${\cal N}$ whose elements are linear
combinations of random $r\times r$ matrices as follows.
\beq
{\cal N} = span\left\{ C,X_2^\ddagger,X_3^\ddagger,\ldots\right\} 
\eeq
Define an inner-product on ${\cal N}$ via
\beq
(A,B) = \ex(\trace[AB\trans]) ,\quad \forall A,B\in {\cal N}
\label{mip}
\eeq
This inner-product is discussed and motivated in \citeN{djwgvar}.
Complete ${\cal N}$ into a Hilbert space, ${\cal M}$. When the space
is completed, limit points such as $HV^\nu H\trans$, $V^\nu$, 
and $F\trans V^\omega F$ are added to the space.
The inner-product on this space is determined by our beliefs
  about the quadratic products, since
\beq
(A,B) = \sum_{j=1}^r\sum_{k=1}^r \left[ \cov(A_{jk},B_{jk}) +
\ex(A_{jk})\ex(B_{jk}) \right]\ \forall A,B\in {\cal M}
\eeq
We may carry out Bayes linear adjustment in this space
 by orthogonal projection of the matrices of interest into subspaces
of observable matrices. Our adjusted expectations for these matrices
 are linear combinations of the prior
matrices and the observable matrices. Note that this 
matrix approach to belief adjustment is a more direct way of getting
at desirable linearity properties of conditional expectations for matrices, 
than via somewhat artificial constructs such as the matrix 
normal distribution.

\subsection{Example}
For our example, we simply construct
\beq
{\cal N} =
span\{C,\bvec{X}^{(1)}_2\bvec{X}^{(1)}_2\trans,\bvec{X}^{(1)}_3
\bvec{X}^{(1)}_3\trans,\ldots, \bvec{X}^{(2)}_3\bvec{X}^{(2)}_3\trans,
\bvec{X}^{(2)}_4\bvec{X}^{(2)}_4\trans,
\ldots\}
\eeq
and impose the inner-product (\ref{mip}), inducing the Hilbert space
${\cal M}$, which contains limit points such as $V^\nu$ and $V^\omega$.
Note that in order to evaluate (\ref{mip}), the 
specifications
 needed are precisely those which were made in Section \ref{exspecs}.
  The fact that many other aspects of the fourth order
specifications are not necessary is very helpful, as this greatly
reduces the specification burden.
Often it is most straightforward to 
 make direct primitive specifications for the matrix object 
inner-product. However, for simplicity in this paper, we have built up
 the specifications for the matrix inner product from
specifications over the scalar quadratic products, 
thus establishing the links between 
the scalar and matrix analysis.
 This is analogous to specifying an expectation of a random quantity by
breaking it up over a partition of events and  specifying 
probabilities over the partition.

\subsection{$n$-step exchangeable matrix objects}
The definition of generalised $n$-step exchangeability applies
directly to 
 matrix objects
in the space ${\cal M}$. The collection of matrix objects
$\{\bvec{X}_t^{{\prime}}\bvec{X}_t^{{\prime}}\trans |\forall
t\geq 2\}$ is
2-step exchangeable in the space ${\cal M}$,
and
the
collection
$\{\bvec{X}_t^{\prime\prime}\bvec{X}_t^{\prime\prime}\trans|\forall
t\geq 3\}$ is 3-step exchangeable. 
This leads to a restatement of Theorem \ref{idquant} for matrices in
the space ${\cal M}$. The limit points are the matrices of limit
points of their elements, due to the consistency of the inner-products
on the scalar and matrix spaces.
\begin{thm}
Put
$
{M}_t =
\frac{1}{2}\left[\bvec{X}^{{\prime}}_t\bvec{X}^{{\prime}}_t\trans-
H\inv(\bvec{X}^{{\prime\prime}}_t\bvec{X}^{{\prime\prime}}_t\trans-
\bvec{X}^{{\prime}}_t\bvec{X}^{{\prime}}_t\trans)H\inv\trans
\right]
$.

\noindent (i) The collection
$
\left\{\left. {M}_t  \right| \forall t\geq 3 \right\}
$
identifies $V^\nu$ in ${\cal M}$.

\noindent (ii) The collection
$
\left\{\left.
 \bvec{X}^{{\prime}}_t\bvec{X}^{{\prime}}_t\trans - M_t -
HM_t H\trans
\right| \forall t\geq 3 \right\}
$
 identifies
$F\trans V^\omega F$ in ${\cal M}$.
\end{thm}
\proofend
Now, for any subspace $D$ such that $C\subseteq D\subseteq{\cal M}$, 
we define {\em adjusted\/} matrix 
expectation, $\ex_D:{\cal M}\rightarrow D$ to be such that $\forall
Y\in
 {\cal M},\ 
\ex_D(Y)$ is the orthogonal projection of $Y$ into $D$, with respect to the 
inner-product $(\cdot,\cdot)$. If $D=span\{C\}$ then we write $\ex(\cdot)$
for $\ex_D(\cdot)$, as this is the usual matrix expectation,
consisting
 of the matrix
of expectations of elements.

\subsection{Adjustment}
Suppose that we are considering observing $n>3$ time points in the
series. Form the matrix space, $\cal M$, and the observable subspace
$D_n\subseteq{\cal M}$
\beq
D_n = span\{C,X_2^\ddagger,X_3^\ddagger,\ldots,X_n^\ddagger\}
\eeq
Then the adjusted expectation map, $\ex_{D_n}(\cdot):{\cal M}
\rightarrow D_n$, is the 
orthogonal projection into the $D_n$ space.
In particular, we evaluate $\ex_{D_n}(V^\nu)$, $\ex_{D_n}(F\trans V^\omega F)$,
which are matrices in the $D_n$ space.
These adjusted expectations are analagous to
posterior expected values for the matrix objects, taking into account
only those limited aspects of the problem which we have considered.
The general relationships between adjusted and posterior beliefs is
discussed in \citeN{mgrevexch}.

\section{Bayes linear adjustment for the example}
\subsection{The adjusted covariance matrices}
Adjustments were carried out using 17 time points from the 
actual time series. The matrix
objects $V^\omega$ and $V^\nu$ were adjusted in the following ways:
\beq
\ex_D(V^\omega) = \left( \begin{array}{rrrrrr}
4.8&0.9&1.0&1.0&0.8&1.5\\
0.9&3.9&1.2&0.9&1.1&0.3\\
1.0&1.2&4.0&1.1&1.1&0.7\\
1.0&0.9&1.1&6.8&0.7&0.8\\
0.8&1.1&1.1&0.7&3.9&0.8\\
1.5&0.3&0.7&0.8&0.8&4.7
\end{array} \right )
\eeq
\beq
\ex_D(V^\nu) = \left( \begin{array}{rrrrrr}
41.8&-5.4&-4.4&-8.0&-4.7&-2.4\\
-5.4&36.7&-3.8&-0.2&-3.2&-4.1\\
-4.4&-3.8&36.1&-4.4&-3.5&-7.5\\
-8.0&-0.2&-4.4&56.6&-5.6&4.8\\
-4.7&-3.2&-3.5&-5.6&34.9&-4.9\\
-2.4&-4.1&-7.5&4.8&-4.9&44.0\\
\end{array} \right )
\eeq
Comparing these with their prior specifications, given in
(\ref{varomega}) and (\ref{varnu}), we see that
the adjusted matrices
are perturbations of the prior expectations for the
matrices.
Notice that the variance associated with the fourth variable has
been inflated considerably in both matrices. The sample variances for
the 17 cases of the six brands we considered were 167, 22, 37, 560, 18 and
427. Informally, it seems that there may indeed be more variability
associated with the fourth (and sixth) variable.

\subsection{First order adjustment}
Since our aim is to predict sales more accurately,
a sensible test of the procedure is to compare the performance
of the first order model, (\ref{exobs}), (\ref{extheta}), using 
both the prior and adjusted
covariance matrices. We find that the Bayes linear diagnostic 
warnings (the {\em size\/} and {\em bearings\/} of the adjustments,
as described in \citeN{mgtraj}) are noticeably closer to their
expected values when
using the adjusted matrices. For the given example, most of the {\em
  size ratios\/} for adjustments of the first order structure were
 noticeably closer to one using the adjusted covariance structure to
predict future values, suggesting that
the adjusted matrices match more closely with the forecast performance
of the model.

\section{Conclusions}
Good forecasting requires careful updating of the covariances within
the time series 
structure. Informally, the degree of {\it shrinkage\/} between the prior
and the data is updated, and relationships between variables are
properly taken into account. Since we are able to adjust the 
covariance matrices for both the {\it observational\/} as well as
{\it state\/} residuals, we are able to properly understand the {\it
competition\/} and {\it demand\/} effects taking place within the
series. By taking a matrix object approach, we greatly simplify the
problem by reducing dimensionality. This is important for both
simplifying belief specification and belief adjustment,
 and also for interpretation
of the structure of the adjustment and accompanying diagnostics.
There are also the general advantages of the Bayes linear approach;
namely of allowing
complete flexibility for the prior specifications, without placing
distributional restrictions on the data or model components.

\section{Acknowledgements}
The first named author is supported by a grant from the UK's EPSRC.
 All Bayes linear computations were carried out using the Bayes linear 
computing
package, \bd described in \citeN{bdworks}, and explained in detail 
in \citeN{gwblincomp}. The covariance
calculations 
for the 
quadratic products were carried 
out using the REDUCE computer algebra system.
The example data was supplied by {\em Positive Concepts Ltd.\/}

\vspace{0.2in}

\begin{appendix}
\noindent{\Large{\bf Appendix}}

\vspace{0.1in}

\noindent{\large{\bf Covariance structure for the example}}

\vspace{0.1in}

\noindent We give here the full covariance structure over the
 quadratic products 
of the one and two step differences, $\bvec{X}_{t}^{{(1)}}$,
$\bvec{X}_{t}^{{(2)}}$ defined by (\ref{xip}), (\ref{xipp}).
 First we need some 
notation for two different ``direct products''
of matrices.
\begin{defn}
For $r\times r$ matrices $A$ (having entry $a_{ij}$ in row $i$, column
$j$) and
$B$ (having entry $b_{ij}$ in row $i$, column
$j$) we define the {\em (left) tensor product\/},
$A\tprod B$ of $A$ and $B$ to be the $r^2\times r^2$ matrix with
the element $a_{jk}b_{lm}$ in row $r(l-1)+j$, column $r(m-1)+k$.
\end{defn}
\begin{defn}
For $r\times r$ matrices $A$ (having entry $a_{ij}$ in row $i$, column
$j$)  and
$B$ (having entry $b_{ij}$ in row $i$, column
$j$)  we define the {\em star product\/} $A\star B$ of $A$ and $B$ to be the
$r^2\times r^2$ matrix with the element $a_{jk}b_{lm}$ in row
$r(l-1)+j$,
column $r(k-1)+m$.
\end{defn}
Given these definitions, the covariance 
structure over the quadratic products of the 1-step
differences is determined by the following relations:
\beq
\cov(\mvec V^\omega,\mvec(\bvec{X}_{t}^{{(1)}}\bvec{X}_{t}^{{(1)}}\trans))
 = \var(\mvec V^\omega)
\eeq
\beq
\cov(\mvec V^\nu,\mvec(\bvec{X}_{t}^{{(1)}}\bvec{X}_{t}^{{(1)}}\trans))
 = 2\var(\mvec V^\nu)
\eeq
\beq
\begin{array}{cccl}
 \cov(\mvec(\bvec{X}_{t}^{{(1)}}\bvec{X}_{t}^{{(1)}}\trans),
  \mvec(\bvec{X}_{t}^{{(1)}}\bvec{X}_{t}^{{(1)}}\trans)) &=&    
&    \var(\mvec V^\omega)+4\var(\mvec V^\nu)\\
&&+&
    \var(\mvec S^\nu_t)
+\var(\mvec S^\nu_{t-1})\\
&&+&
    \var(\mvec S^\omega_{t}) + 
      2[ \ex(V^\nu\tprod V^\nu)+\ex(V^\nu\star V^\nu) ]\\
  &&+ &4[ \ex(V^\nu)\tprod\ex(V^\omega)+
 \ex(V^\nu)\star \ex(V^\omega)\\
&&+& \ex(V^\omega)\tprod \ex(V^\nu)+
 \ex(V^\omega)\star \ex(V^\nu)
 ]\\
\end{array}
\eeq
\beq
\cov(\mvec(\bvec{X}_{t}^{{(1)}}\bvec{X}_{t}^{{(1)}}\trans),
  \mvec(\bvec{X}_{t-1}^{{(1)}}\bvec{X}_{t-1}^{{(1)}}\trans)) =
  4(\var(\mvec V^\nu)+\var(\mvec V^\omega)) +
  \var(\mvec S^\nu_{t-1})
\eeq
\beq
\cov(\mvec(\bvec{X}_{t}^{{(1)}}\bvec{X}_{t}^{{(1)}}\trans),
  \mvec(\bvec{X}_{t-s}^{{(1)}}\bvec{X}_{t-s}^{{(1)}}\trans)) =
  4\var(\mvec V^\nu)+\var(\mvec V^\omega) \quad\forall t,\forall s\geq 2
\eeq
The covariance structure over the quadratic products of the 2-step
differences are given below.
\beq
\cov(\mvec V^\omega,\mvec(\bvec{X}_{t}^{(2)}\bvec{X}_{t}^{(2)}\trans))
 = 2\var(\mvec V^\omega)
\eeq
\beq
\cov(\mvec V^\nu,\mvec(\bvec{X}_{t}^{(2)}\bvec{X}_{t}^{(2)}\trans))
 = 2\var(\mvec V^\nu)
\eeq
\beq
\begin{array}{cccl}
 \cov(\mvec(\bvec{X}_{t}^{(2)}\bvec{X}_{t}^{(2)}\trans),
  \mvec(\bvec{X}_{t}^{(2)}\bvec{X}_{t}^{(2)}\trans)) &=&    
&    4\var(\mvec V^\omega)+4\var(\mvec V^\nu)
+
    \var(\mvec S^\nu_{t})\\
&&
+&\var(\mvec S^\nu_{t-2})
+
    \var(\mvec S^\omega_{t})+\var(\mvec S^\omega_{t-1})\\
  &&+ &2[ \ex(V^\nu\tprod V^\nu)+\ex(V^\nu\star V^\nu)+
\ex(V^\omega\tprod V^\omega)
\\
&&+&
      \ex(V^\omega\star V^\omega) ]+ 4[ \ex(V^\nu)\tprod\ex(V^\omega)+
 \ex(V^\nu)\star\ex(V^\omega)\\
&&+& \ex(V^\omega)\tprod\ex(V^\nu)+
 \ex(V^\omega)\star\ex(V^\nu)
 ]\\
\end{array}
\eeq
\beq
\cov(\mvec(\bvec{X}_{t}^{(2)}\bvec{X}_{t}^{(2)}\trans),
  \mvec(\bvec{X}_{t-1}^{(2)}\bvec{X}_{t-1}^{(2)}\trans)) =
  4[\var(\mvec V^\nu)+\var(\mvec V^\omega)] +
  \var(\mvec S^\omega_{t-1})
\eeq
\beq
\cov(\mvec(\bvec{X}_{t}^{(2)}\bvec{X}_{t}^{(2)}\trans),
  \mvec(\bvec{X}_{t-2}^{(2)}\bvec{X}_{t-2}^{(2)}\trans)) =
  4[\var(\mvec V^\nu)+\var(\mvec V^\omega)] +
  \var(\mvec S^\nu_{t-2})
\eeq
\beq
\cov(\mvec(\bvec{X}_{t}^{(2)}\bvec{X}_{t}^{(2)}\trans),
  \mvec(\bvec{X}_{t-s}^{(2)}\bvec{X}_{t-s}^{(2)}\trans)) =
  4\var(\mvec V^\nu)+\var(\mvec V^\omega) \quad\forall
 t,\forall s\geq 3
\eeq
The covariances between the one and two step differences are
determined as follows:
\beq
\cov(\mvec(\bvec{X}_{t}^{{(1)}}\bvec{X}_{t}^{{(1)}}\trans),
  \mvec(\bvec{X}_{t+s}^{(2)}\bvec{X}_{t+s}^{(2)}\trans)) =
  4\var(\mvec V^\nu)+2\var(\mvec V^\omega) \quad\forall t,\forall s\geq 3
\eeq
\beq
\cov(\mvec(\bvec{X}_{t}^{{(1)}}\bvec{X}_{t}^{{(1)}}\trans),
  \mvec(\bvec{X}_{t+2}^{(2)}\bvec{X}_{t+2}^{(2)}\trans)) =
  4\var(\mvec V^\nu)+2\var(\mvec V^\omega) +
  \var(\mvec S_{t-2})
\eeq
\beq
\begin{array}{cccl}
 \cov(\mvec(\bvec{X}_{t}^{{(1)}}\bvec{X}_{t}^{{(1)}}\trans),
  \mvec(\bvec{X}_{t+1}^{(2)}\bvec{X}_{t+1}^{(2)}\trans)) &=&    
&    2\var(\mvec V^\omega)+4\var(\mvec V^\nu)\\
&&+&
    \var(\mvec S^\nu_{t-2})+
    \var(\mvec S^\omega_{t-1})\\
  &&+ & \ex(V^\nu)\tprod\ex(V^\omega)+
 \ex(V^\nu)\star\ex(V^\omega)\\
&&+& \ex(V^\omega)\tprod\ex(V^\nu)+
 \ex(V^\omega)\star\ex(V^\nu)\\
\end{array}
\eeq
\beq
\begin{array}{cccl}
 \cov(\mvec(\bvec{X}_{t}^{{(1)}}\bvec{X}_{t}^{{(1)}}\trans),
  \mvec(\bvec{X}_{t}^{(2)}\bvec{X}_{t
}^{(2)}\trans)) &=&    
&    2\var(\mvec V^\omega)+4\var(\mvec V^\nu)\\
&&+&
    \var(\mvec S^\nu_{t})+
    \var(\mvec S^\omega_{t})\\
  &&+ & \ex(V^\nu)\tprod\ex(V^\omega)+
 \ex(V^\nu)\star\ex(V^\omega)\\
&&+& \ex(V^\omega)\tprod\ex(V^\nu)+
 \ex(V^\omega)\star\ex(V^\nu)\\
\end{array}
\eeq
\beq
\cov(\mvec(\bvec{X}_t^{{(1)}}\bvec{X}_{t}^{{(1)}}\trans),
  \mvec(\bvec{X}_{t-1}^{(2)}\bvec{X}_{t-1}^{(2)}\trans)) =
  4\var(\mvec V^\nu)+2\var(\mvec V^\omega) +
  \var(\mvec S^\nu_{t-1})
\eeq
\beq
\cov(\mvec(\bvec{X}_{t}^{{(1)}}\bvec{X}_{t}^{{(1)}}\trans),
  \mvec(\bvec{X}_{t-s}^{(2)}\bvec{X}_{t-s}^{(2)}\trans)) =
  4\var(\mvec V^\nu)+2\var(\mvec V^\omega) \quad\forall
  t,\forall s\geq 2
\eeq

These results are obtained by focussing on a general element of a matrix on the
left hand side, and then substituting into the left hand sides 
 the definitions 
(\ref{xip}) and (\ref{xipp}), 
expanding the
covariances, substituting representations (\ref{omegarep}) and 
(\ref{nurep}), and then 
simplifying the result using known
orthogonalities to deduce the general element of the matrices on the
right hand side. However,
there are several hundred terms in some of the expansions and a 
computer algebra
package was used to ensure the accuracy of the results.

\end{appendix}

\newpage

\bibliographystyle{chicago}
\bibliography{../../bib/bayeslin,../../bib/djw}

\end{document}